\documentclass{emulateapj}
\usepackage[T1]{fontenc}
\usepackage[latin9]{inputenc}
\usepackage{amssymb}
\usepackage{graphicx}
\usepackage{natbib}
\usepackage{psfrag}

\newcommand{\noun}[1]{\textsc{#1}}

\shorttitle{Synoptic Survey Discovered AM CVn System}
\shortauthors{Levitan et al.}

\begin{document}

\title{PTF1\,J071912.13+485834.0: An outbursting AM CVn system discovered
by a synoptic survey}

\author{David Levitan\altaffilmark{1}, Benjamin J. Fulton\altaffilmark{2},
Paul J. Groot\altaffilmark{3,4}, Shrinivas R. Kulkarni\altaffilmark{4},
Eran O. Ofek\altaffilmark{4,5}, Thomas A. Prince\altaffilmark{1},
Avi Shporer\altaffilmark{2,6}, Joshua S. Bloom\altaffilmark{7},
S. Bradley Cenko\altaffilmark{7}, Mansi M. Kasliwal\altaffilmark{4},
Nicholas M. Law\altaffilmark{8}, Peter E. Nugent\altaffilmark{9},
Dovi Poznanski\altaffilmark{5,7,9}, Robert M. Quimby\altaffilmark{4},
Assaf Horesh\altaffilmark{4}, Branimir Sesar\altaffilmark{4}, and Assaf Sternberg\altaffilmark{10} }

\altaffiltext{1}{Department of Physics, California Institute of Technology, Pasadena, CA 91125, USA}\altaffiltext{2}{Las Cumbres Observatory Global Telescope Network, 6740 Cortona Drive, Suite 102, Santa Barbara, CA 93117, USA }\altaffiltext{3}{Department of Astrophysics IMAPP, Radboud University Nijmegen, PO Box 9010, NL-6500 GL Nijmegen, the Netherlands}\altaffiltext{4}{Cahill Center for Astrophysics, California Institute of Technology, Pasadena, CA 91125, USA}\altaffiltext{5}{Einstein fellow}\altaffiltext{6}{Department of Physics, Broida Hall, University of California, Santa Barbara, CA 93106, USA}\altaffiltext{7}{Department of Astronomy, University of California, Berkeley, CA 94720-3411, USA}\altaffiltext{8}{Dunlap Institute for Astronomy and Astrophysics, University of Toronto, 50 St. George Street, Toronto M5S 3H4, Ontario, Canada}\altaffiltext{9}{Computational Cosmology Center, Lawrence Berkeley National Laboratory, 1 Cyclotron Road, Berkeley, CA 94720, USA}\altaffiltext{10}{Benoziyo Center for Astrophysics, Faculty of Physics, Weizmann Institute of Science, Rehovot 76100, Israel}

\begin{abstract}
We present extensive photometric and spectroscopic observations of
PTF1\,J071912.13+485834.0, an outbursting AM CVn system discovered by
the Palomar Transient Factory (PTF). AM CVn systems are stellar binaries with some
of the smallest separations known and orbital periods ranging from 5 to 65 minutes.
They are believed to be composed of a white dwarf accretor and a (semi)-degenerate
He-rich donor and are considered to be the helium equivalents of
Cataclysmic Variables. We have spectroscopically and photometrically
identified an orbital period of $26.77\pm0.02$ minutes for PTF1\,J071912.13+485834.0
and found a super-outburst recurrence time of greater than 65 days
along with the presence of ``normal'' outbursts --- rarely seen
in AM CVn systems but well known in super-outbursting Cataclysmic
Variables. We present a long-term light curve over two super-cycles
as well as high cadence photometry of both outburst and quiescent
stages, both of which show clear variability. We also compare both
the outburst and quiescent spectra of PTF1\,J071912.13+485834.0 to
other known AM CVn systems, and use the quiescent phase-resolved
spectroscopy to determine the origin of the photometric variability.
Finally, we draw parallels between the
different subclasses of SU UMa-type Cataclysmic Variables and outbursting
AM CVn systems. We conclude by predicting that the Palomar Transient
Factory may more than double the number of outbursting AM CVn systems
known, which would greatly increase our understanding of AM CVn systems.
\end{abstract}


\section{Introduction}

AM CVn systems --- ultra-compact semi-detached binaries --- are stellar binaries with some of the smallest separations known.
They have been found with orbital periods ranging from 5 to 65 minutes.
The prototype, AM CVn, was initially identified
as a possible binary star by \citet{1967AcA....17..255S} and was eventually
theorized to be composed of a relatively massive white dwarf accretor
and a much lower mass semi-degenerate or degenerate helium-transferring
donor \citep{1967AcA....17..287P,1972ApJ...175L..79F}. AM CVn systems are believed
to be one of the strongest Galactic low-frequency gravitational wave sources \citep{2004MNRAS.349..181N,2007MNRAS.382..685R}
and the source of the proposed ``.Ia'' supernovae \citep{2007ApJ...662L..95B}.
We refer the reader to \citet{2005ASPC..330...27N} and \citet{2010PASP..122.1133S}
for reviews.

Short period systems --- those with orbital periods below roughly 20
minutes --- are in a constant state of high mass transfer from the secondary
to the optically thick accretion disk. They are known as ``high''
state systems, and their spectra, dominated by the accretion disk,
are characterized by broad, shallow helium absorption lines with few other features.
High state systems have been observed to have superhumps --- photometric
variability of $\sim0.1\,\text{mag}$ with a period slightly longer than the orbital period \citep[e.g.][]{1997PASP..109.1100P}
--- similar to those those found in SU UMa-type Cataclysmic Variables \citep{1996PASP..108...39O}.

At the other end of the period range are the quiescent systems
with orbital periods above roughly 40 minutes. They are believed to have
low mass transfer rates and an optically thin disk. Instead of absorption
lines, these systems have prominent helium emission lines in their
spectra. Quiescent systems do not show prominent photometric variability.

Between these two period ranges are the so-called ``outbursting''
AM CVn systems, which feature outbursts similar to those found in
dwarf novae-type Cataclysmic Variables. While in the ``high''
state, these systems exhibit the properties of short-period AM CVn
systems, and while in the ``quiescent'' state, they exhibit properties
of the long-period AM CVn systems \citep[see e.g.][]{2007MNRAS.379..176R}. In outburst they are typically
3-5 magnitudes brighter than in quiescence and feature superhumps.
These outbursts tend to last for a few weeks, and recur on a timescale (where known)
between 46 days \citep[e.g.][]{2000MNRAS.315..140K} and over a year \citep[e.g.][]{2011MNRAS.410.1113C}.
Between the two states, some of these systems have been observed to have a
``cycling'' state wherein some experience magnitude changes of
$\sim1\,\text{mag}$ with a period of about a day \citep{2000PASP..112..625P}.
One system, CR Boo, has also been found to have {}``normal'' outbursts
that last 1-2 days and recur every 4-8 days \citep{2000MNRAS.315..140K}, as
opposed to the longer ``super-outbursts'' described previously.

AM CVn systems have been extensively compared to Cataclysmic Variables
(CVs). Of primary interest for this comparison are the SU UMa-type dwarf novae-type CVs, which
exhibit both super-outbursts
and normal outbursts, but with somewhat longer typical recurrence
times than outbursting AM CVn systems. See \citet{WARNER1995}
for an extensive review. While the
photometric behavior is similar between dwarf novae and AM CVn systems,
the chemical composition, structure of the donor, and evolutionary
pathways are very different.

Only 26 AM CVn systems have been reported in the literature. Being intrinsically rare
and with colors similar to those of ordinary white dwarfs, they are difficult to discover
and population estimates have proven to be difficult to calculate
\citep[e.g.][]{2001A&A...368..939N,2007MNRAS.382..685R}. 
Initially, AM CVn systems were serendipitous discoveries, typically as a result
of their photometric variability or color. More recently, the population has almost doubled as a 
result of the Sloan Digital Sky Survey (SDSS).
Seven systems were discovered from a search for spectra containing helium emission lines
\citep{2005MNRAS.361..487R,2005AJ....130.2230A,2008AJ....135.2108A} and five more from
a follow-up color selection and spectroscopic survey \citep{2009MNRAS.394..367R,2010ApJ...708..456R}.

However, the wide variety of photometric variability exhibited by AM CVn systems
makes them an effective target for large-scale, synoptic surveys.
The most recent published new AM CVn
system, with an orbital period of 15.6 minutes, was discovered in
{\it Kepler} satellite data from its superhump-induced photometric variability \citep{2011ApJ...726...92F}.
Here, we present a new AM CVn system discovered in outburst by the Palomar Transient
Factory --- the first system discovered by a systematic, synoptic survey covering thousands of square degrees.

The Palomar Transient Factory\footnote{\url{http://www.astro.caltech.edu/ptf}} (PTF) uses the Oschin 48-inch
telescope (P48) at the Palomar Observatory
to image $7.2\deg^2$ with each exposure. In a typical night, up to $\sim2,000\deg^{2}$ are observed to a depth of $R\sim20.6$
\citep{2009PASP..121.1395L,2009PASP..121.1334R}. 

We begin by describing the discovery of PTF1\,J071912.13+485834.0\footnote{``PTF1''
refers to preliminary versions of the PTF catalog, as opposed to sources from the final catalog, which will use ``PTF''. 
It is possible that a source in the PTF1 catalog will have slightly different coordinates in the PTF catalog.} (hereafter
PTF1J0719+4858) and summarizing our follow-up observations. In \S\ref{sec:observations_photometric}
we present photometric observations.
We describe the features of both outburst and quiescent spectra in
\S\ref{sec:observations_spectra}, as well as the determination of
the spectroscopic period from phase-resolved spectroscopy. In \S\ref{sec:Discussion}, we compare
PTF1J0719+4858 to other outbursting AM CVn systems,
discuss the source of the quiescent photometric variability,
and consider how many more such systems can be discovered by PTF.
Finally, we summarize in \S\ref{sec:Summary}.

\section{Discovery and Summary of Observations}

\label{sec:Detection-and-Classification}
\begin{figure}
\begin{centering}
\includegraphics{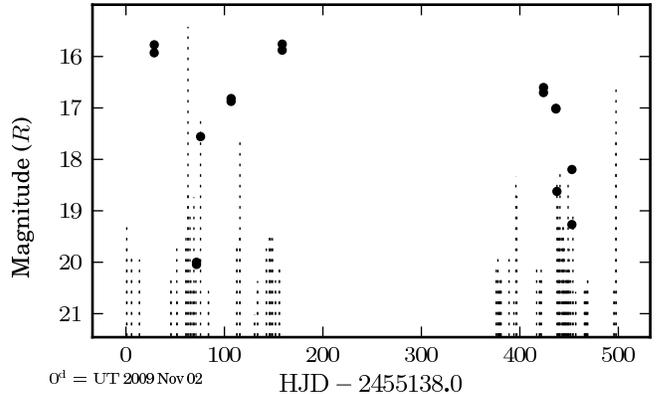}
\end{centering}
\caption{PTF light curve of PTF1J0719+4858. Discovery occurred at the first
data point. Magnitudes were obtained by difference
imaging relative to a deep co-add reference image as part of the PTF Transient
Pipeline. The tops of the dashed lines represent limits for non-detections and are
theoretical calculations based on observed conditions. Errors on observations
in outburst are $\sim0.01\,\text{mag}$ and errors in quiescence are
$\sim0.1\,\text{mag}$. Note that these observations are in $R$ band
as opposed to the $g'$ band used in the rest of this paper. PTF1J0719+4858
is a blue object and hence fainter in $R$. }
\label{fig:ptflc}
\end{figure}

PTF1J0719+4858 was detected
in outburst by the Palomar Transient Factory at $R=15.8$ on 2009
December 01 and classified as a transient with the designation
PTF09hpk\footnote{This is a PTF transient designation. For stellar discoveries, we use
the more conventional IAU variable star coordinate name.}.
A graphical summary of the PTF photometry can be found in Figure \ref{fig:ptflc}.

A classification spectrum was taken using Keck-I/LRIS \citep{1998SPIE.3355...81M}
on 2010 January 14 and reduced using standard \noun{iraf} tasks. Noticing
the lack of a redshift, the PTF extragalactic team classified the
spectrum as a Cataclysmic Variable. In a subsequent inspection of the
PTF spectral database, we noticed the presence of multiple distinct,
double-peaked helium emission lines, some with a central peak (we
refer the reader to \S\ref{sub:High-Speed-Spectroscopy} which
contains a high signal-to-noise quiescence spectrum), and it was re-classified
as an AM CVn system candidate.

We focused our follow-up efforts on both long-term monitoring and
short time-scale variability studies, using the Palomar $60''$ telescope
(P60) and two telescopes from the Las Cumbres Observatory Global Telescope
Network \citep[LCOGT; ][]{2010arXiv1011.6394S}: the 2-m Faulkes Telescope
North (FTN) and the $32''$ Byrne Observatory at Sedgwick (BOS). Between
October 2010 and March 2011, we obtained a total of 195 exposures
in good weather for our long-term photometric monitoring campaign
using P60 and FTN with a goal of obtaining at least one exposure
per night. We present this light curve in \S\ref{sub:long-term-photometric}.
We also observed PTF1J0719+4858 at high cadence several times to characterize
the photometric variability on the order of the orbital period, both in quiescence
and in outburst. We discuss the periods identified
from these observations in \S\ref{sub:high-cadence-photometry}.

Besides photometric observations, we also obtained individual spectra
of PTF1J0719+4858 on multiple nights, as well as phase-resolved spectroscopy.
Individual spectra of PTF1J0719+4858 were obtained with Keck-I, the
William Herschel Telescope, and the Palomar $200''$ Hale Telescope
in October and November, 2010.  To obtain the orbital period,
we obtained roughly four hours of spectroscopic observations with Keck-I/LRIS using
three minute exposures. These observations are presented in \S\ref{sub:High-Speed-Spectroscopy}.

\section{Photometric Observations and Results}

\label{sec:observations_photometric}

\subsection{Analysis and Reduction Process}

Palomar $60''$ data was de-biased and flat-fielded using the P60
pipeline \citep{2006PASP..118.1396C}. The FTN data was processed using the LCOGT
pipeline. The BOS data was de-biased and flat-fielded using \noun{iraf}
tasks, astrometrically calibrated using \noun{Astrometry.net} \citep{2010AJ....139.1782L}, and
cosmic rays were removed using the \noun{L.A. Cosmic} algorithm
\citep{2001PASP..113.1420V}. The \noun{Sextractor} package \citep{1996A&AS..117..393B} was
used to identify sources in each exposure and their instrumental magnitudes
were obtained using optimal point spread function photometry \citep{1998MNRAS.296..339N} 
as implemented by the \noun{Starlink}\footnote{The \noun{Starlink} Software Group homepage can be
found at http://starlink.jach.hawaii.edu/starlink.} package \noun{autophotom}.

Light curves were calculated using a matrix-based, least squares minimization, relative photometry algorithm.
The primary goal of any such algorithm is to minimize
noise, typically by assuming certain stars in the field are non-variable
and identifying an optimal zero point for the exposure. We expanded on
this to simultaneously solve for both the zero-point and additional
de-trending terms that corrected for airmass and instrument changes.
The algorithm is similar to that developed in \citet{1992PASP..104..435H}
and is described in the appendix of \citet{2011arXiv1103.3010O}.

To accomplish the de-trending, we modeled each observation as
\begin{displaymath}
m_{i,j}=\overline{M}_{j}+Z_{i}+\alpha c_{j}A_{i}+\sum_{k=1}^{n_k}\beta_{k}c_{j}
\end{displaymath}
where the needed data is:
\begin{itemize}
\item $m_{i,j}$: the magnitude of source $j$ on exposure $i$.
\item $c_{j}$: a color for each source. The color is required to
compensate for the stronger effects of airmass on blue stars, as well
as the the differences in CCD efficiency over a range of wavelengths.
For our light curves, we used $c_{j}=g'_{j}-r'_{j}$, where $g'_{j}$ and $r'_{j}$
refer to the magnitudes of the $j$th source in the respective SDSS filters.
\item $A_{i}$: the airmass of each exposure.
\end{itemize}
and the terms to be fitted are:
\begin{itemize}
\item $Z_{i}$: the optimal zero-point term of each exposure.
\item $\overline{M}_{j}$: the mean magnitude term of the source.
\item $\alpha$: the airmass calibration coefficient for all exposures and sources
\item $\beta_{k}$: the $k$th telescope/instrument calibration coefficient, for $k=1,2,\ldots, n_k$ where $n_k$ is the number of telescopes. This term is introduced to take into account the different responses of each telescope/instrument. For light curves with data from only one instrument, these terms were not used.
\end{itemize}
It is important to ensure that all stars used for the solution (``calibration
stars'') are themselves not variable. We restricted the stars used to those
found in 80--100\% of exposures, depending on the light curve, and
iteratively removed any sources with high residuals. Since the solution
is not unique unless reference magnitudes are provided, we used blue
magnitudes from USNO-B 1.0.

This algorithm provided very good results --- even light curves taken
over months with different telescopes and conditions obtained a magnitude
scatter (RMS) of $\sim0.035\,\text{mag}$ for $g'\approx16$ and $\sim0.055\,\text{mag}$
for $g'\approx19.4$, the quiescent magnitude of PTF1J0719+4858. The
RMS errors provided with the figures in this paper are based on the
median scatter of other stars with similar magnitude present in at
least 50\% of observations. Additionally, individual errors --- the
combination of the Poisson error and the fit errors --- are provided
for some of the light curves. These are typically very close to the
magnitude scatter, except for those exposures obtained during bad weather.

For high cadence light curves, period analysis was performed with
\noun{SigSpec} \citep{2007A&A...467.1353R}. All default options were used,
except as noted for individual cases, and weights for measurements
were always provided. \noun{SigSpec} produces a list of significant
periods and corresponding ``sig'' values. A ``sig'' value
of $c$ means that the period has a chance of 1 in $10^{c}$ of being
noise.

\subsection{Long-term Photometric Behavior}

\label{sub:long-term-photometric}
\begin{figure*}
\begin{centering}
\includegraphics{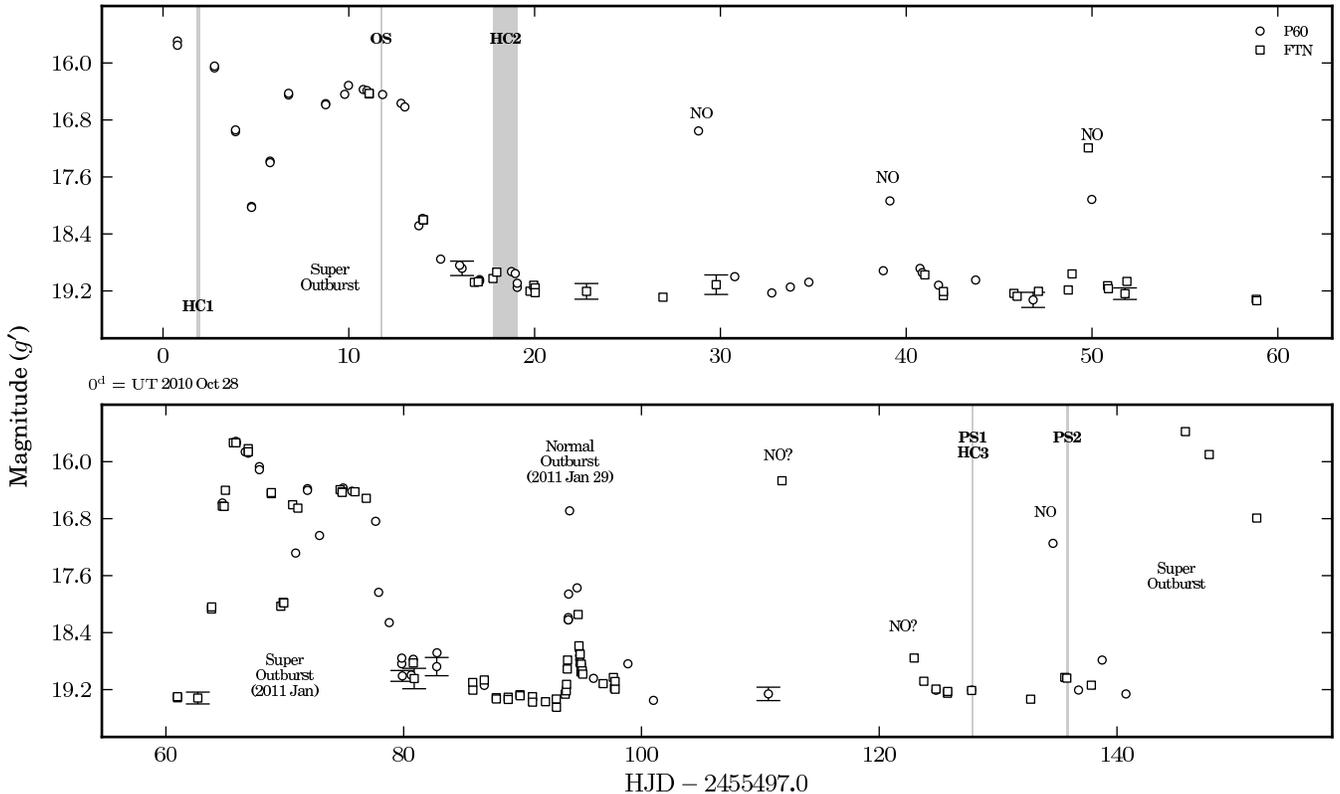}
\par\end{centering}

\caption{Long term light curve of PTF1J0719+4858 with 90 exposures from P60
and 105 exposures from FTN. Exposures from both telescopes were 60\,s and
we attempted to obtain 1--2 exposures per night.
29 calibration stars were used to achieve an RMS of $\sim0.055\,\text{mag}$
in quiescence and an RMS of $\sim0.035\,\text{mag}$ in outburst (small enough
that the error bars cannot be seen). Observations with $\sigma>0.075\,\text{mag}$
(due to weather or other issues) are marked with error bars. Note the difference in
the scale of the time axis for the upper and lower panels: the
top half shows the first super-outburst cycle ($\sim 60\,$d) and the bottom half
shows the second cycle ($\sim 80\,$d).\\
The  2011 January super-outburst (the only one observed in its entirety) is labeled as
is the normal outburst of 2011 January 29. The other data points
we believe to be normal outbursts are labeled as \emph{NO}.
Two of these points are questionable due to lack of observations,
and are identified with a question mark. The quiescent magnitude is $g'\approx19.4$.\\
The high cadence runs in Table \ref{fig:High-Cadence-Photometry}
are identified as HC$n$ while the phase-resolved spectroscopy runs
discussed in \S\ref{sub:High-Speed-Spectroscopy} are identified
as PS$n$. The outburst spectrum discussed in \S\ref{sub:indiv-spectra}
is labeled as OS. \label{fig:Long-Term-Photometry}}
\end{figure*}

The long term light curve of PTF1J0719+4858 from FTN and P60 is presented
in Figure \ref{fig:Long-Term-Photometry}. We note the pattern of ``high'' states and ``quiescent''
states. Additionally, we note the presence of ``normal'' outbursts
(as opposed to the ``super-outbursts'' more commonly associated
with AM CVn systems).

We observed the January 2011 super-outburst in its entirety and find
a rise time from the last measurement in quiescence to the peak magnitude
of $3.2$ days with $\Delta\text{mag}=3.6$. Immediately following
this rise, we see a drop to a plateau that may be the cycling state
seen in other AM CVn systems \citep[e.g.][]{2000PASP..112..625P}. Finally,
22 days after the beginning of the super-outburst, PTF1J0719+4858
returned to quiescence.

The recurrence time was significantly different between the two super-cycles
we observed. We approximate (assuming the behavior of the super-outburst
itself is the same) that the recurrence time from the first super-outburst
to the second was 65 days. However, the recurrence time from the second
super-outburst to the third is greater than 78 days (this uncertainty
is due to weather impacting our observations).

\begin{figure}
\begin{centering}
\includegraphics{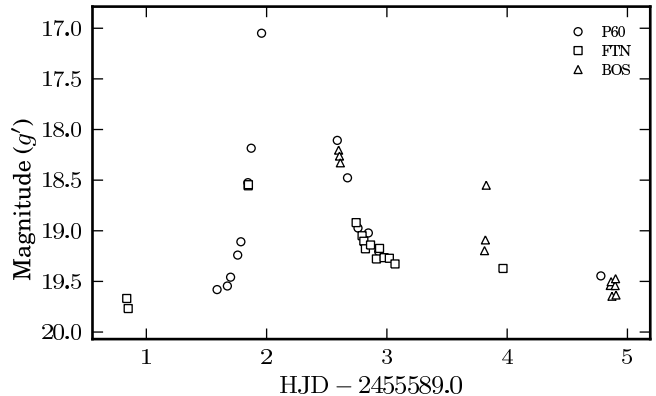}
\end{centering}

\caption{Normal outburst of PTF1J0719+4858 (2011 Jan 29). The light curve contains
41 exposures, with 14 from P60 (60\,s), 15 from FTN (60\,s), and 12 from
BOS (300\,s). 45 calibration stars were used. At $g'\approx17$, $\text{RMS}\approx0.02\,\text{mag}$
and at $g'\approx19.5$, $\text{RMS}\approx0.04\,\text{mag}$.}
\label{fig:normal-outburst}
\end{figure}

Between super-outbursts, we observed normal outbursts in PTF1J0719+4858,
which have also been identified in CR Boo \citep{2000MNRAS.315..140K}. After
initially observing single data points indicating a sudden jump in
luminosity, we successfully predicted the 2011 January 29 outburst
and obtained a total of 41 exposures, of which 12 are from BOS and
are not included in the long term light curve. The light curve of
this outburst is presented in Figure \ref{fig:normal-outburst}. 
 During this normal outburst, PTF1J0719+4858 experienced a luminosity
increase of 2.5 magnitudes over 7 hours from the last observation
at $g'>19.4$ to the brightest point measured. If we consider only
the increase from $g'>19$ we find an increase of 2.1 magnitudes over
only 4 hours.

The recurrence time of the normal outbursts was stable throughout the first super-cycle.
We observed three normal outbursts with a recurrence time of $\sim10.5$ days. This most likely
represents all the normal outbursts in this super-cycle, due to our almost daily coverage. The delay
between the end of the second super-cycle and the first normal outburst was the same as that in
the first super-cycle. However, it appears that the normal outburst recurrence time was significantly
different in the second super-cycle, likely associated with the longer recurrence time of the super-outburst.
Given our poor coverage (as a result of weather), we cannot make any statements about these differences.

\subsection{High-Cadence Photometry}

\label{sub:high-cadence-photometry}Short-term photometric variability was detected in several high cadence
observations of PTF1J0719+4858 (see Table \ref{fig:High-Cadence-Photometry} and labels in Figure \ref{fig:Long-Term-Photometry}).

\begin{deluxetable}{ccccc}
\tablecaption{High cadence photometry runs of PTF1J0719+4858\label{fig:High-Cadence-Photometry}}
\tablecolumns{5}
\tablehead{\colhead{Label} & \colhead{Telescope} & \colhead{UT Date(s)} & \colhead{Exposures} & \colhead{Read-Out}\\
& & & & \colhead{Time (s)}}
\tablecomments{All exposures taken with a $g'$ filter.}
\startdata
HC1 & P60 & 2010 Oct 29 & $92 \times 45$\,s & $10$\tablenotemark{a}\\
HC2 & BOS & 2010 Nov 14/15 & $54/60 \times 300$\,s & $31$\\
HC3 & P60 & 2011 Mar 4 &$ 87 \times 60$\,s & $24$
\enddata
\tablenotetext{a}{Taken in quarter-chip mode, which decreased read-out time}
\end{deluxetable}

\begin{figure}
\includegraphics{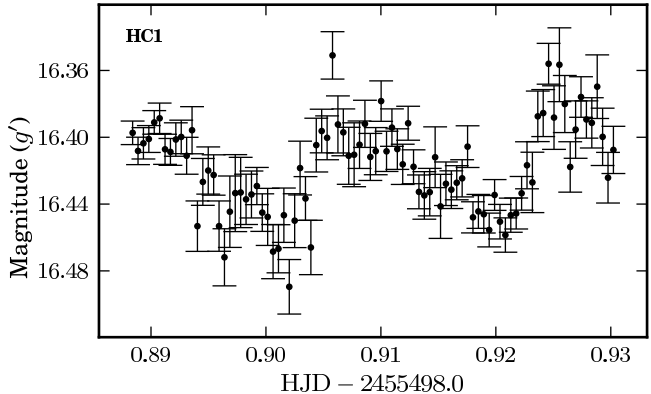}

\caption{Superhumps of PTF1J0719+4858. The light curve was constructed using
12 calibration stars, and other sources at this magnitude had an RMS
of $\sim0.015\,\text{mag}$. The shape is consistent with superhumps in
similar systems such as CR Boo \citep{1997PASP..109.1100P} and KL Dra \citep{2002MNRAS.334...87W}.}
\label{fig:superhumps}
\end{figure}

During the first super-outburst (labeled HC1), we detected photometric variability of
$\Delta\text{mag}\approx0.1$ (see Figure \ref{fig:superhumps}) with a shape consistent with that of superhumps in
other AM CVn systems \citep[e.g.][]{1997PASP..109.1100P,2002MNRAS.334...87W}. Analysis of the period using a Lomb-Scargle periodogram
suggests it is $\sim27$ minutes, while a \noun{SigSpec} analysis found a period of $\sim26$ minutes.
Given that we observed only two cycles, we cannot state a more accurate superhump period.

\begin{figure}
\begin{centering}
\includegraphics{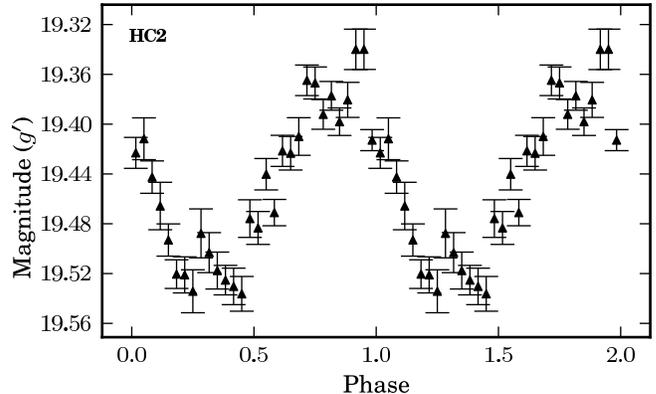}
\end{centering}
\caption{Phase-binned light curve of PTF1J0712+4858 in quiescence from
114 BOS exposures of 300\,s folded on a period of $1606.3$\,s. 18 calibration stars were used to construct
the light curve. At this magnitude, $\text{RMS}\approx0.035\,\text{mag}$.
Error bars for individual points are based on the standard deviation of all
measurements in that phase bin.
An arbitrary zero phase of $\text{HJD}=2455514.831555$ (the start of the
observations) was used.}
\label{fig:bos-qui-lc}
\end{figure}

A second set of high-cadence
observations (labeled HC2) was obtained almost immediately after PTF1J0719+4858 returned to quiescence following
the first observed super-outburst. Here, we see photometric variability of $\Delta\text{mag}\approx 0.2$.
We performed a \noun{SigSpec} analysis of the light curve for these two nights with the AntiAIC anti-aliasing feature enabled
and found a period of $1606.3\pm2.5$\,s with a ``sig'' of $13.7$. Since we observed many periods of this variability, we 
present a phase-binned light curve in Figure \ref{fig:bos-qui-lc}.  Additional observations that allow
a more precise determination of the superhump period could be used along with the quiescent photometric period to determine
a mass ratio \citep{2005PASP..117.1204P}.

We also obtained a set of high cadence observations (labeled HC3) coincident with our phase-resolved
spectroscopy at Keck-I (see  \S\ref{sub:High-Speed-Spectroscopy}). 
These observations also showed photometric variability,with
$\Delta\,\text{mag}\approx0.06$. PTF1J0719+4858 was in quiescence at the time of the observations.
A \noun{SigSpec} analysis identified a period of $1550\,$s with a sig of 3.1. Given the short
observation time and the low significance, we folded the light curve
at both this period and the spectroscopic period obtained in \S\ref{sub:High-Speed-Spectroscopy}.
These produced similar results and, thus, we believe that the true
period is that obtained via phase-resolved spectroscopy. We discuss the
simultaneous photometry and spectroscopy in \S\ref{sub:simult-spec-photom}.

\section{Spectroscopic Observations and Results}

\label{sec:observations_spectra}

\subsection{Follow-up Spectra}

\label{sub:indiv-spectra}
\begin{figure*}
\begin{centering}
\includegraphics{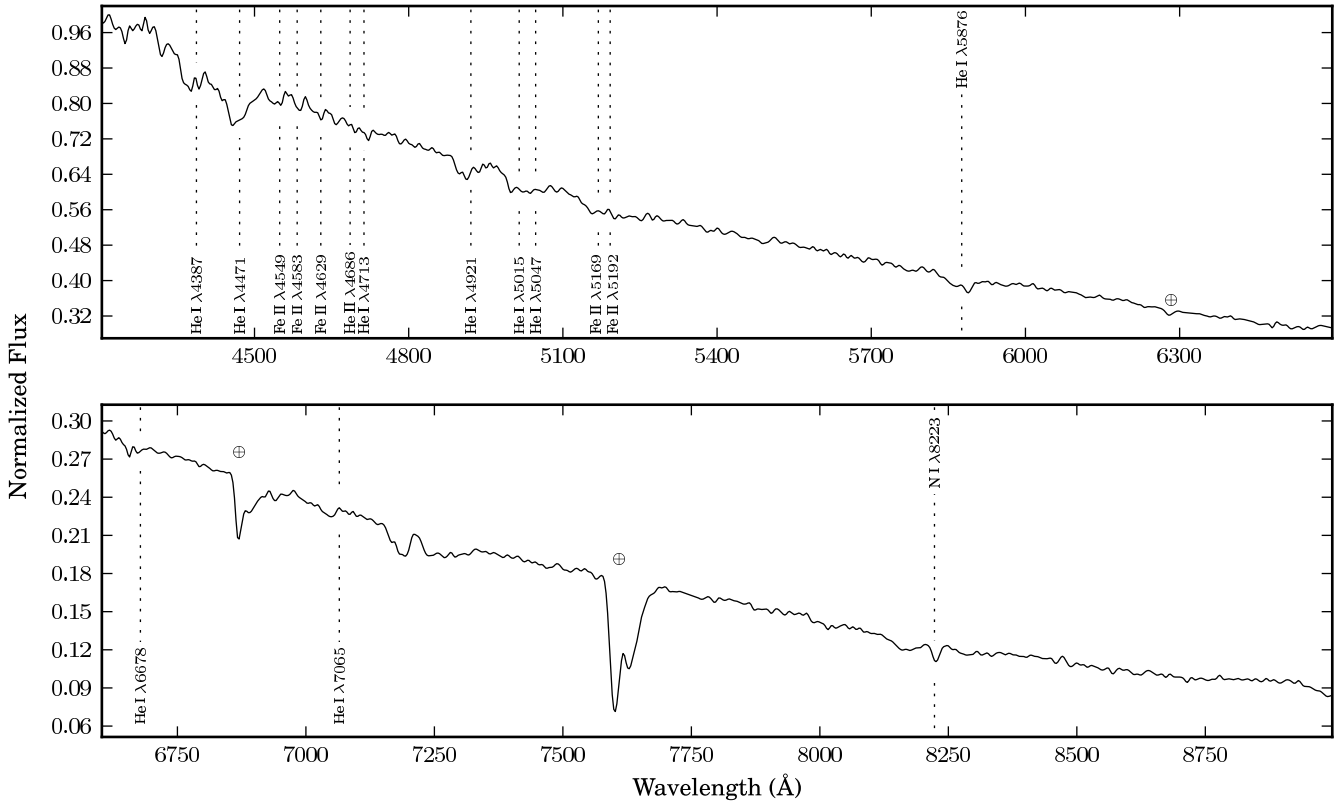}
\par\end{centering}

\caption{Outburst spectrum of PTF1J0719+4858 taken with WHT/ACAM on 2010 November
6. Strong helium absorption lines are present throughout
the spectrum, as well a \ion{Fe}{2} lines. We also highlight the \ion{N}{1} $\lambda8223$
absorption line, which has not been seen before in an AM CVn system in the
high state.}
\label{fig:outburst-spec}
\end{figure*}

The identification spectra were reduced as
part of the PTF spectroscopic program using standard \noun{iraf} tasks.
We present a typical outburst spectrum --- taken with WHT/ACAM \citep{2008SPIE.7014E.229B}
on 2010 November 6 and labeled as OS in Figure \ref{fig:Long-Term-Photometry}
--- in Figure \ref{fig:outburst-spec} with the prominent lines identified.
 The outburst spectra varied slightly, with absorption lines being
more prominent on some than on others. However, \ion{He}{1} $\lambda4471$ was always
visible. The best spectrum in quiescence is the co-added spectrum
of the phase-resolved observations, shown in Figure \ref{fig:coadd-spectra},
again with lines identified.

The quiescence spectrum features very broad, double-peaked
emission lines, some with a possible central spike. Shortward of $4000\,$\AA,
the spectrum shows an interplay of lines that is consistent
with \ion{Ca}{2} H \& K emission interwoven with \ion{He}{1} $\lambda3888$ \citep[see][]{2006MNRAS.365.1109R},
but the low resolution of the current observation and the broadness
of the lines makes their presence difficult to establish. We can establish the
presence of \ion{Fe}{2}, Si, and \ion{N}{1} emission lines in the rest of the spectrum.
It is also possible that \ion{He}{2} $\lambda4200$ is seen in the spectrum, albeit
very weak. \ion{He}{2} $\lambda4200$ has not been previously
seen in an AM CVn system. In the outburst spectrum, we see weak absorption
lines of \ion{He}{1} and \ion{He}{2}, as well as \ion{Fe}{2}. Si is not seen, but we note
the presence of \ion{N}{1} $\lambda8223$, which has not been seen this strong in other
high-state AM CVn systems.

Table \ref{fig:equivalent-widths} lists the equivalent widths of the most prominent emission lines.
Based on the presence of the noted elements, we find that that the
spectra of PTF1J0719+4858 are most similar to those of 2003aw \citep{2006MNRAS.365.1109R}
and SDSSJ0804+1616 \citep{2009MNRAS.394..367R}. Future work in identifying
abundances may shed light on the chemical composition and evolutionary history of such systems \citep{2010MNRAS.401.1347N}.

\begin{deluxetable}{cc}
\tablecaption{Equivalent widths of identified lines from co-added quiescence spectrum\label{fig:equivalent-widths}
}
\tablecolumns{2}
\tablehead{\colhead{Line} & \colhead{Equivalent Width (\AA)}}
\startdata
\ion{He}{1} $\lambda4387$ & $-4.9\pm0.1$\\
\ion{He}{1} $\lambda4471$ & $-9.4\pm0.1$\\
\ion{He}{2} $\lambda4686$ + \ion{He}{1} $\lambda4713$ & $-9.8\pm0.3$\\
\ion{He}{1} $\lambda4921$ & $-3.7\pm0.1$\\
\ion{He}{1} $\lambda5015 + 5047$ & $-5.2\pm0.2$\\
\ion{He}{1} $\lambda5876$ & $-14.7\pm0.3$\\
\ion{He}{1} $\lambda6678$ & $-10.7\pm0.4$\\
\ion{He}{1} $\lambda7065$ & $-7.2\pm0.4$
\enddata
\end{deluxetable}

\subsection{High Speed Spectroscopy in Quiescence}

\label{sub:High-Speed-Spectroscopy}

\begin{figure*}
\begin{centering}
\includegraphics{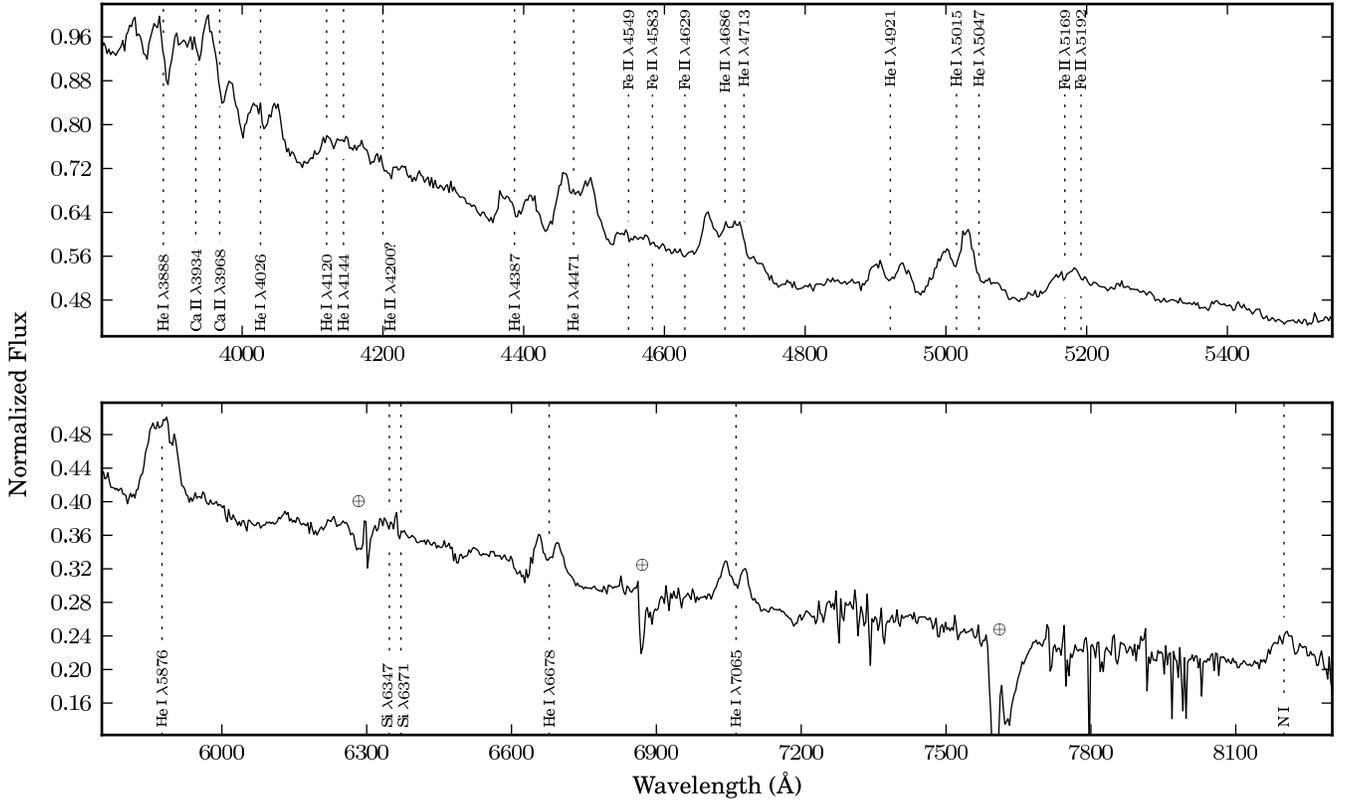}
\par\end{centering}

\caption{Co-added Keck-I/LRIS quiescence spectrum of PTF1J0719+4858. A total
of 72 exposures of 180\,s were co-added for the blue side. For the red
side, 62 exposures of 180\,s were median co-added to remove cosmic ray
effects. Strong helium lines are evident, along with \ion{Fe}{2} and Si $\lambda6347+6371$.
A broad emission feature of \ion{N}{1} is also present at $\sim8200\,$\AA.}
\label{fig:coadd-spectra}
\end{figure*}

Here, we discuss the phase resolved spectroscopy undertaken at the Keck Observatory (see Table
\ref{fig:Phase-Resolved-Spectroscopy}). The spectra were reduced using optimal extraction \citep{1986PASP...98..609H}
as implemented in the \noun{Pamela} code \citep{1989PASP..101.1032M} as well
as the \noun{Starlink} packages \noun{kappa}, \noun{figaro}, and \noun{convert}.
For these exposures, wavelength calibration exposures were taken at
the beginning, middle, and end of each set of observations using the
Hg, Cd, and Zn lamps for the blue CCD and the Ne and Ar lamps for
the red CCD. Wavelength calibration for individual spectra was interpolated
between these calibration spectra. We present a co-added spectrum
in Figure \ref{fig:coadd-spectra}.

\begin{deluxetable}{cccccc}
\tablecaption{Phase-Resolved Spectroscopy using Keck-I/LRIS\label{fig:Phase-Resolved-Spectroscopy}
}
\tablecolumns{6}
\tablenotetext{a}{Spectral $\times$ Spatial}
\tablehead{\colhead{UT Date} & \colhead{CCD} & \colhead{Disp. Elem.} & \colhead{Bins\tablenotemark{a}} & \colhead{Slit (\arcsec)} & \colhead{Exposures}}
\startdata
2011 Mar 04 & Blue & 600/4000 & $4\times4$ & $1.5$ & $37 \times 180$\,s\\
2011 Mar 04 & Red & 600/7500 & $4\times4$ & $1.5 $& $35 \times 180$\,s\\
2011 Mar 12 & Blue & 600/4000 & $4\times2$ & $0.7$ & $32 \times 180$\,s\\
2011 Mar 12 & Red & 600/7500 & $4\times2$ & $0.7$ & $30 \times 180$\,s
\enddata
\tablecaption{Binning was increased to reduce read-out noise and time. The conditions were substantially better
on 2011 March 12, allowing the use of less binning the spatial direction.}
\end{deluxetable}

\begin{figure}
\includegraphics{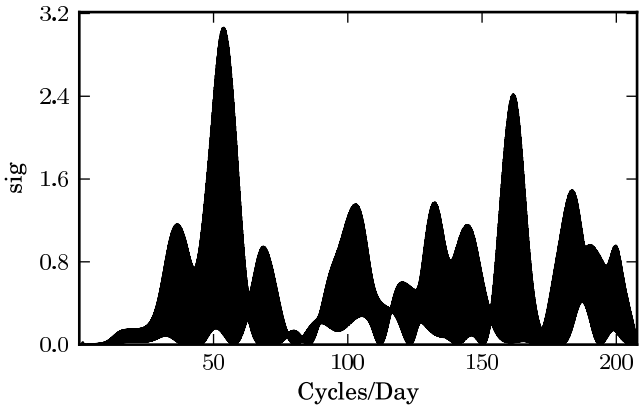}

\caption{Signficance spectrum of helium emission line flux ratios obtained
from Keck-I/LRIS phase-resolved spectroscopy on 2011 March 04/12 (generated
by \noun{SigSpec}).}
\label{fig:flux-ratio-sigspec}
\end{figure}

We use a technique similar to previous analyses of AM CVn system phase-resolved
spectra, first developed by \citet{1981ApJ...244..269N}, to establish the
orbital period. Each spectrum was rebinned to the same wavelengths
and the location of individual emission lines were identified. Because
of the large number of cosmic rays on the red side despite processing
with \noun{L.A. Cosmic} \citep{2001PASP..113.1420V}, we concentrated on the blue
side and used the helium lines at $4026\,$\AA, $4387\,$\AA,
$4471\,$\AA, $4686\,$\AA, $4921\,$\AA, and
$5015\,$\AA. The lines from each exposure were re-binned and
co-added to produce a summed He emission line. We then subtracted
the red 40\% of the line from the blue 40\% of the line, and divided
by the continuum. This produced a time series of flux ratios, which
we analyzed using \noun{SigSpec}. We identified a period of $1606.2\pm0.9$\,s
 with a ``sig'' of 3.1. The statistical significance spectrum
is in Figure \ref{fig:flux-ratio-sigspec}.

\begin{figure}
\begin{centering}
\includegraphics{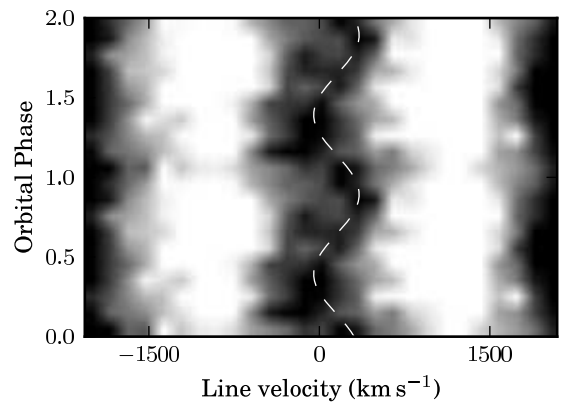}
\end{centering}

\caption{Phase-folded, trailed spectrum of the combined He emission lines visible
on Keck-I/LRIS's blue CCD. The period is set to $1606.2$\,s. The
``S-wave'', marked by a dashed line, is faint but present. Variation in gray scale indicates
relative flux densities. The two white columns are the double-peaked
structure of the He emission lines. The start of the first night of
observations, at $\text{HJD}=2455624.81212$, was used as the zero phase.}

\label{fig:s-wave}
\end{figure}

While not a very high confidence level, we believe the above period is, in fact,
the orbital period, for two reasons. First, the period found is within
the error bars of the previously discussed quiescent photometric period
(see \S\ref{sub:high-cadence-photometry}). Second, the movement
of the disk's hotspot can be identified by creating a phase-binned,
trailed spectrum of the He emission line. The rotation of the disk
produces an {}``S-wave'' that has been observed previously in known
systems \citep[e.g.][]{2005MNRAS.361..487R,2010ApJ...708..456R}. In
the case of PTF1J0719+4858, this S-wave is very weak, but still discernible.
We present the trailed spectrum in Figure \ref{fig:s-wave}.
 This is not the first system where the S-wave was difficult to detect.
As reported in \citet{2009MNRAS.394..367R}, an S-wave in SDSSJ0804+1616 was
not found in one of two sets of spectra. Further observations are
required to establish whether PTF1J0719+4858 has similar variability
in the strength of the S-wave or if it is simply a weak feature.

\section{Discussion}

\label{sec:Discussion}

\subsection{Comparison of long-term light curve with that of other AM CVn systems}

We broadly group the known outbursting AM CVn systems into two
categories: those having super-outbursts that occur at least every three months
and those with less frequent super-outbursts. The latter group has either poorly
determined or undetermined recurrence times. We summarize
their properties in Table \ref{fig:outbursting-amcvns}.

\begin{deluxetable*}{ccccc}
\tablecaption{Properties of known outbursting AM CVn systems\label{fig:outbursting-amcvns}}
\tablecolumns{5}
\tablehead{\colhead{Object} & \colhead{Orbital Period (s)} & \colhead{Super-outburst} & \colhead{$\Delta\text{mag}$} & \colhead{References}\\
& & \colhead{Recurrence Time} & &}
\tablenotetext{a}{Reported as 46.3\,d in \citet{2000MNRAS.315..140K}, but reported as variable in \citet{2001IBVS.5120....1K} based on additional data}
\tablenotetext{b}{Superhump period}
\tablenotetext{c}{These systems have no published recurrence time, but it is believed to be significantly longer than 3 months}
\startdata
CR Boo & 1471 & 46.3\,d\tablenotemark{a} & 4.5 & \citet{2000MNRAS.315..140K,2001IBVS.5120....1K} \\
KL Dra & 1500 & \,50\,d & 4.2 & \citet{2002MNRAS.334...87W}, Levitan et al., in prep\\
V803 Cen & 1596 & \,77\,d & 4.6 & \citet{2004PASJ...56L..39N,2007MNRAS.379..176R}\\
PTF1J0719+4858 & 1606 & 65\text{--}80\,d & 3.5 & this paper\\
\\
SDSS J0926+3624 & 1699 & 104--449\,d & 3.3 &  \citet{2011MNRAS.410.1113C}\\
CP Eri & 1701 & \nodata$\!\!\!$\tablenotemark{c} & 3.2 & \citet{1992ApJ...399..680A,2001ApJ...558L.123G}\\
2003aw & 2028 &\nodata$\!\!\!$\tablenotemark{c} & 4.8 & \citet{2004PASJ...56L..39N,2006MNRAS.365.1109R}\\
2QZ J1427-01 & \,\,\,2194\tablenotemark{b} & \nodata$\!\!\!$\tablenotemark{c} & 5.3 & \citet{2005IAUC.8531....3W}\\
SDSS J1240-0159 & 2241 & \nodata$\!\!\!$\tablenotemark{c} & 4.5 & \citet{2005MNRAS.361..487R,Woudt1240,2011arXiv1104.0107S}\\
SDSS J0129+3842 & \,\,\,2274\tablenotemark{b} & \nodata$\!\!\!$\tablenotemark{c} & 4.6 &  \citet{2005AJ....130.2230A,2011arXiv1104.0107S}\\
SDSS J2047+0008 & \nodata & \nodata$\!\!\!$\tablenotemark{c} & $\sim 6$ & \citet{2008AJ....135.2108A}
\enddata
\end{deluxetable*}

It appears that these sources can be cleanly divided by orbital
period, with the break between 1606\,s and 1699\,s. For the first
group, the recurrence times are fairly well determined and appear to correlate with the orbital
period. The second group is more difficult to understand, primarily due to a lack of known recurrence times.
The one published recurrence time, for SDSSJ0926, is very poorly determined \citep{2011MNRAS.410.1113C}.
However, there does appear to be a clear gap between the determined
recurrence times of PTF1J0719+4858 and much more poorly determined recurrence time
of SDSSJ0926+3624, and we question whether this difference in recurrence time
is purely a result of the increased orbital period (and thus decreased
mass transfer rates) or if different parameters also place a role
(such as the mass of the primary and/or the entropy of the secondary).

We can draw parallels to the much more common Cataclysmic
Variables (CVs). The class of CVs most like outbursting AM CVn systems are the
SU UMa-type CVs, which also exhibit both normal outbursts and super-outbursts
with superhumps \citep{WARNER1995}. SU UMa-type systems
typically have super-outburst recurrence times of several hundred days and
orbital periods between 90 and 120 minutes \citep{1996PASP..108...39O}, but
have been found to have two extreme cases. The ER UMa-type systems have
recurrence times between 19 and 45 days \citep{2002A&A...386..891B}, while
WZ Sge-type systems have recurrence times of decades and lower orbital
periods of 80--90 minutes. The long recurrence times of WZ Sge-type systems
are explained by the lower mass transfer rates of WZ Sge-type systems,
as a result of their more evolved state \citep[CVs are believed to have
negative $\dot{P}$, as opposed to AM CVn systems; ][]{1996PASP..108...39O}.

In \citet{2000MNRAS.315..140K}, CR Boo was proposed to be the helium equivalent
of an ER UMa-type Cataclysmic Variable. However, given that all four of
the frequently outbursting systems appear to have fairly similar behavior,
we believe that this is typical behavior for AM CVn systems with
these orbital periods, as opposed to an extreme case. On the other
hand, the recurrence times of the longer period outbursting AM CVn
systems indicate that these systems are more akin to the WZ Sge-type CVs.
This is consistent with the assumption that AM CVn systems have positive
$\dot{P}$ and thus have lower mass transfer with greater orbital
periods. Additional discoveries of AM CVn systems, as well as better,
more systematic observations of known systems, are necessary to understand
the reality of the difference in recurrence times.

\subsection{Origin of photometric variability in quiescence}
\label{sub:simult-spec-photom}
\begin{figure*}
\begin{centering}
\includegraphics{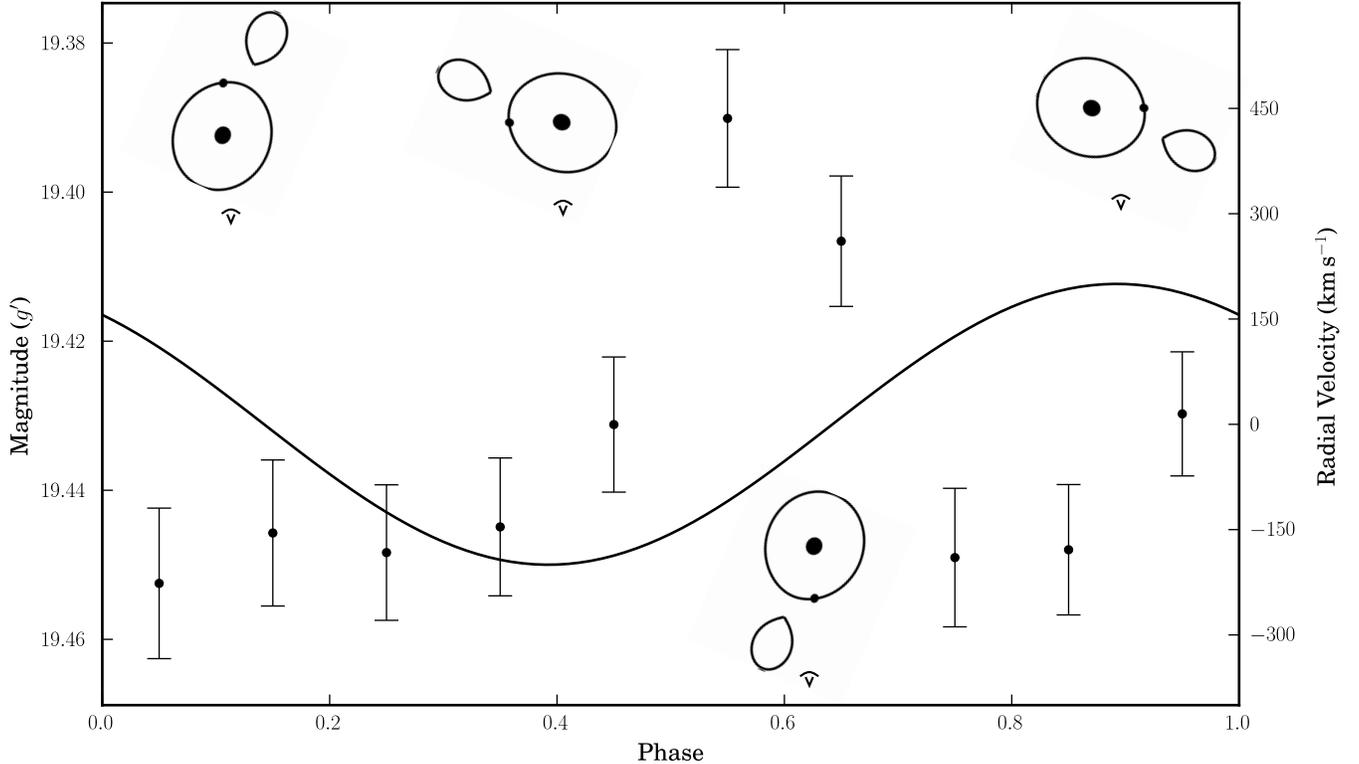}
\end{centering}

\caption{Phase-binned light curve taken on 2011 Mar 04 using the P60. 87 60\,s exposures
were binned into 10 bins with a zero phase of $\text{HJD}=2455624.81212$, the start
of the phase-resolve spectroscopy, and the spectroscopically derived period of $26.77\,$m.
Overplotted is the S-wave found in \S\ref{sub:High-Speed-Spectroscopy}.
The rough orientation of the binary is shown with drawings at the four extreme points of the orbit,
along with the position of the observer.\\
We see that the increase in brightness is coincident with the radial velocity of the hot spot being
roughly zero and the time period preceding it, indicating that the hot spot and the previously-heated
disk edge is the likely source of the increased brightness.}

\label{fig:simult-spec-photom}
\end{figure*}

The high-cadence photometric observations discussed in \S\ref{sub:high-cadence-photometry}
and labeled as HC3 were coincident with the phase-resolved spectroscopy discussed in \S\ref{sub:High-Speed-Spectroscopy},
providing us with an opportunity to determine the source of the observed photometric variability.
We present a binned, phase-folded photometric light curve in Figure \ref{fig:simult-spec-photom}, together 
with the radial velocity of the hot spot. The binning provides an increase in the signal-to-noise
and gives a roughly $5\sigma$ detection. We remind the reader that HC3 was observed in
quiescence, during which time superhumps are not believed to occur in either AM CVn systems  or CVs.

The increase in brightness immediately follows
the blue-shifted portion of the radial velocity curve and is
coincident with a lack of radial velocity. This indicates that the
photometric variability in PTF1J0719+4858 is caused by
the hot spot and the associated heated edge of the disk
being closest to the observer (see drawings in Figure \ref{fig:simult-spec-photom}).
We assume here that the S-wave is, in fact, caused by the hot spot.

AM CVn system variability in quiescence has been observed for
other non-eclipsing systems such as CR Boo \citep{1997ApJ...480..383P}
and KL Dra \citep{2002MNRAS.334...87W}, but no study has been done
of the origin. However, such studies for CVs have linked
the hot spot to the observed photometric variability \citep[e.g.][]{1983A&A...128...37S}.

The concern with this explanation is the lack of precision
in the photometric data and the lack of time resolution as a result of the binning.
This also makes it difficult to compare Figure \ref{fig:simult-spec-photom} to
observations of other AM CVn systems or CVs. However, we believe that there
is sufficient data to link the quiescent variability to the location of the hotspot relative
to the observer.

\subsection{Potential for future discoveries with PTF}

How many additional AM CVn systems might
be discovered by PTF? First, we estimate the fraction of all AM CVn systems
that are outbursting. Consider a simple model of a single system's evolution
by assuming that gravitational wave radiation and
angular momentum loss from mass transfer are solely responsible
for the evolution of the orbital period \citep{1967AcA....17..287P,2001A&A...368..939N}. Then, using typical mass values for
an AM CVn system at its minimum period ($0.6 M_\odot$ \& $0.25 M_\odot$, although similar values do not significantly affect the results)
we find that an AM CVn system spends $\sim0.5\%$ of its life between orbital periods of 20 and 27 minutes (the frequent outbursters)
and $\sim3.3\%$ of of its life between orbital periods of 27 and 40 minutes (the less frequent outbursters). We use these
numbers as a simple estimate of the percentage of AM CVn systems that are outbursting.

We now estimate the number of systems PTF could discover. Given that AM CVn
systems have outbursts with $\Delta\text{mag}\gtrsim3$, we conservatively assume
that any outbursting system with a quiescent magnitude of $\lesssim23$ can be detected
in outburst by PTF. We assume a scale height of 300\,pc \citep{2007MNRAS.382..685R} and
a scale length of 2.5\,kpc \citep{1997ApJ...483..103S} in the Galaxy and an AM CVn system space density
of $\rho_{o}=3.1\times10^{-6}\,\text{pc}^{-3}$ \citep[the observed space density based on pessimistic population models from][]{2007MNRAS.382..685R}.
Given that the systems are in quiescence, we further assume that the accretor provides all of the system's luminosity, and use eq. (5) from
\citet{2007MNRAS.382..685R}, which is a parametrization of Figure 2 in \citet{2006ApJ...640..466B},
to calculate the absolute magnitudes of the AM CVn systems. This likely
means that our estimate is conservative since the disk is known to provide
part of the luminosity \citep[$\sim30\%$ for SDSSJ0926+3624; see][]{2007ASPC..372..431M} in quiescence.
We find that there are approximately $\frac{1.3\,\text{systems}}{100\deg^{2}}$ with orbital
periods between 20 and 40 minutes and at $20^{\circ}<|b|<60^{\circ}$ (the galactic latitudes for most of PTF's observations).
Given PTF's footprint of $10,000\deg^{2}$, we estimate that PTF might detect up to 136 such systems. However, the
uncertainty in the AM CVn system population density estimates
likely means that this number is only accurate to within a factor of 10.

However, the apparent long outburst recurrence times of longer period
systems will make these much more difficult to detect. If we consider only those
systems with frequent outbursts and thus orbital periods between 20
and 27 minutes, we find that there are up to 18 such systems. Given the recurrence times of 45--80 days,
the presence of normal outbursts in at least two systems, and the super-outburst duty cycle
of 30--50\%, it is very likely that all 18 systems can be detected as part of the
PTF transient search.

We note that searches in lower galactic latitudes make detection
much more likely. The population distribution of outbursting AM CVn systems
almost doubles between $20^{\circ}<|b|<25^{\circ}$ to $15^{\circ}<|b|<20^{\circ}$. Although
the analysis of lower-latitude data is more difficult due to the larger number of sources
overall, it still presents the best opportunity to discover a large number of AM CVn systems.

Despite the exciting possibilities
of using synoptic surveys to search for outbursting systems with quiescent
magnitudes up to $R\sim23\text{--}25$ (and even deeper for future surveys),
we caution that confirmation of these systems will be difficult. The established
method for finding orbital periods is via phase-resolved spectroscopy,
requiring large telescopes and short exposure times for even fairly
bright objects. Even objects with a quiescent magnitude of $R\sim23$ cannot
be observed in such a fashion with today's telescopes. Instead,
such systems can be observed in outburst. The hot spot
has been observed in high state AM CVn systems \citep{2006MNRAS.371.1231R}
and it is likely that it can be seen in the high state of outbursting
systems as well. Additionally, as demonstrated with PTF1J0719+4858
and other AM CVn systems, photometric periods can be obtained from
both superhumps and (potentially) in quiescence, providing a good
estimate of the orbital period.

\section{Summary}

\label{sec:Summary}

We have presented extensive photometric and spectroscopic observations
of PTF1\,J071912.13+485834.0. We have observed the system in both quiescence
and outburst, observing the strong emission lines and a weak photometric
period in the former, and absorption lines and detectable superhumps
in the latter. From the phase-resolved spectroscopy, we have identified
a weak, albeit detectable, signal in the spectrum that indicates an
orbital period of $1606.2\pm0.9$\,s. This data has, in combination with the
simultaneous high-cadence photometry, allowed us to determine
the possible source of the quiescent photometric variability. We have also looked at
the long-term light curve and found a variable super-outburst recurrence
time, as well as regularly occurring normal outbursts. Based on the
identified spectroscopic period, the double peaked emission lines
in quiescence, and its photometric behavior, we classify PTF1J0719+4858
as an AM CVn system.

PTF1J0719+4858 has the longest orbital period of known, frequently
outbursting AM CVn systems. We have calculated that PTF has the capability
to significantly increase the number of such systems and potentially find
many more systems with less regular outbursts. Additional discoveries
would expand our understanding of both the structure and evolution of AM CVn systems and
their population density.

\acknowledgements

Observations obtained with the Samuel Oschin Telescope at the Palomar
Observatory as part of the Palomar Transient Factory project, a scientific
collaboration between the California Institute of Technology, Columbia
University, Las Cumbres Observatory, the Lawrence Berkeley National
Laboratory, the National Energy Research Scientific Computing Center,
the University of Oxford, and the Weizmann Institute of Science. Some
of the data presented herein were obtained at the W.M. Keck Observatory,
which is operated as a scientific partnership among the California
Institute of Technology, the University of California and the National
Aeronautics and Space Administration. The Observatory was made possible
by the generous financial support of the W.M. Keck Foundation. The
William Herschel Telescope is operated on the island of La Palma by
the Isaac Newton Group in the Spanish Observatorio del Roque de los
Muchachos of the Instituto de Astrofísica de Canarias. This paper
uses observations obtained with facilities of the Las Cumbres Observatory
Global Telescope. The Byrne Observatory at Sedgwick (BOS) is operated
by the Las Cumbres Observatory Global Telescope Network and is located
at the Sedgwick Reserve, a part of the University of California Natural
Reserve System.

S.B.C.~wishes to acknowledge generous support from Gary and
Cynthia Bengier, the Richard and Rhoda Goldman Fund, National Aeronautics and
Space Administration (NASA)/{\it Swift} grant NNX10AI21G, NASA/{\it Fermi} grant
NNX1OA057G, and National Science Foundation (NSF) grant AST--0908886.

\textit{Facilities:}
\facility{PO:1.2m}, \facility{PO:1.5m}, \facility{LCOGT}, \facility{BOS},
\facility{FTN (Spectral)}, \facility{Hale (DBSP)}, \facility{ING:Herschel (ACAM)}, \facility{Keck:I (LRIS)}\\

\bibliographystyle{apj}
\bibliography{ms}

\end{document}